\documentclass{article}
\usepackage{amsmath,graphicx,mlspconf,color,subcaption,amssymb,nicefrac,bm}
\usepackage[subtle]{savetrees} 
 
\title{Cognitive-driven convolutional beamforming using \\ EEG-based auditory attention decoding}
\name{Anonymous\thanks{Anonymous.}}
\address{Anonymous}
\name{%
    Ali Aroudi$^{\dagger\star}$%
    \quad Marc Delcroix$^{\dagger}$%
    \quad Tomohiro Nakatani$^{\dagger}$%
    \quad Keisuke Kinoshita$^{\dagger}$%
    \quad Shoko Araki$^{\dagger}$%
    \quad Simon Doclo$^{\star}$
}
\address{%
    $^{\dagger}$ NTT Communication Science Laboratories, NTT Corporation, Kyoto, Japan \\
    $^{\star}$ Department of Medical Physics and Acoustics and Cluster of Excellence Hearing4All,\\ University of Oldenburg, Oldenburg, Germany 
}

\begin{document}
\ninept

\maketitle

\begin{abstract}
The performance of speech enhancement algorithms in a multi-speaker scenario depends on correctly identifying the target speaker to be enhanced. Auditory attention decoding (AAD) methods allow to identify the target speaker which the listener is attending to from single-trial EEG recordings.  Aiming at enhancing the target speaker and suppressing interfering speakers, reverberation and ambient noise, in this paper we propose a cognitive-driven multi-microphone speech enhancement system, which combines a neural-network-based mask estimator,  weighted minimum power distortionless response convolutional beamformers and AAD. 
To control the suppression of the interfering speaker, we also propose an extension incorporating an interference suppression constraint. The experimental results show that the proposed system outperforms the state-of-the-art cognitive-driven speech enhancement systems in challenging reverberant and noisy conditions.
\end{abstract}
%
%
\section{Introduction}
\label{sec:intro}
In a multi-speaker scenario the  performance  of many speech  enhancement  algorithms depends on correctly identifying the target speaker to be enhanced. Recent advances in electroencephalography (EEG) have shown  that  it  is  possible  to identify the target speaker which the listener is attending to using single-trial EEG-based auditory attention decoding (AAD) methods \cite{OSullivan2014, Ali_2019_neural_system, Alickovic_2019, Gregory_2019}.  However, many AAD methods rely on the unrealistic assumption that the clean speech signals of the speakers are available as reference signals for decoding. In real-world conditions, obviously only the microphone signals, which consist of a mixture of the speakers, including reverberation and background noise, are available.

Aiming at incorporating AAD in speech enhancement, several algorithms have recently been proposed to generate appropriate reference signals for decoding from the microphone signals  \cite{Han_2019, Simon_Eyndhoven_2016, Tao_2019_ICASSP, Aroudi_2020_cognitive-driven_beamforming_ASLP}. Most cognitive-driven speech enhancement algorithms generate reference signals by separating the speakers from the mixture received at the microphones either using time-domain neural networks \cite{Han_2019}, multi-channel Wiener filters \cite{Simon_Eyndhoven_2016} or minimum variance distortionless response (MVDR) beamformers \cite{Aroudi_2020_cognitive-driven_beamforming_ASLP}. Using AAD, one of the reference signals is then selected as the enhanced attended speaker. More recently, aiming  at  controlling  the  suppression  of the  interfering  speaker, which  is  important  when  intending  to  switch  attention  between speakers,  a cognitive-driven beamforming system using linearly  constrained  minimum  variance (LCMV) beamformers has been proposed \cite{Tao_2019_ICASSP, Aroudi_2020_cognitive-driven_beamforming_ASLP}.  

While most aforementioned cognitive-driven speech enhancement systems are able to suppress the interfering speakers and background noise, they may not be able to suppress (late) reverberation,  which is known to have a detrimental effect on speech quality and intelligibility \cite{Warzybok_2013}. In this paper we propose a cognitive-driven convolutional beamforming system aiming at enhancing the attended speaker and jointly suppressing the interfering speakers, reverberation and background noise. 

\begin{figure}[t]
 \centering
  \centerline{\includegraphics[width=8.5cm]{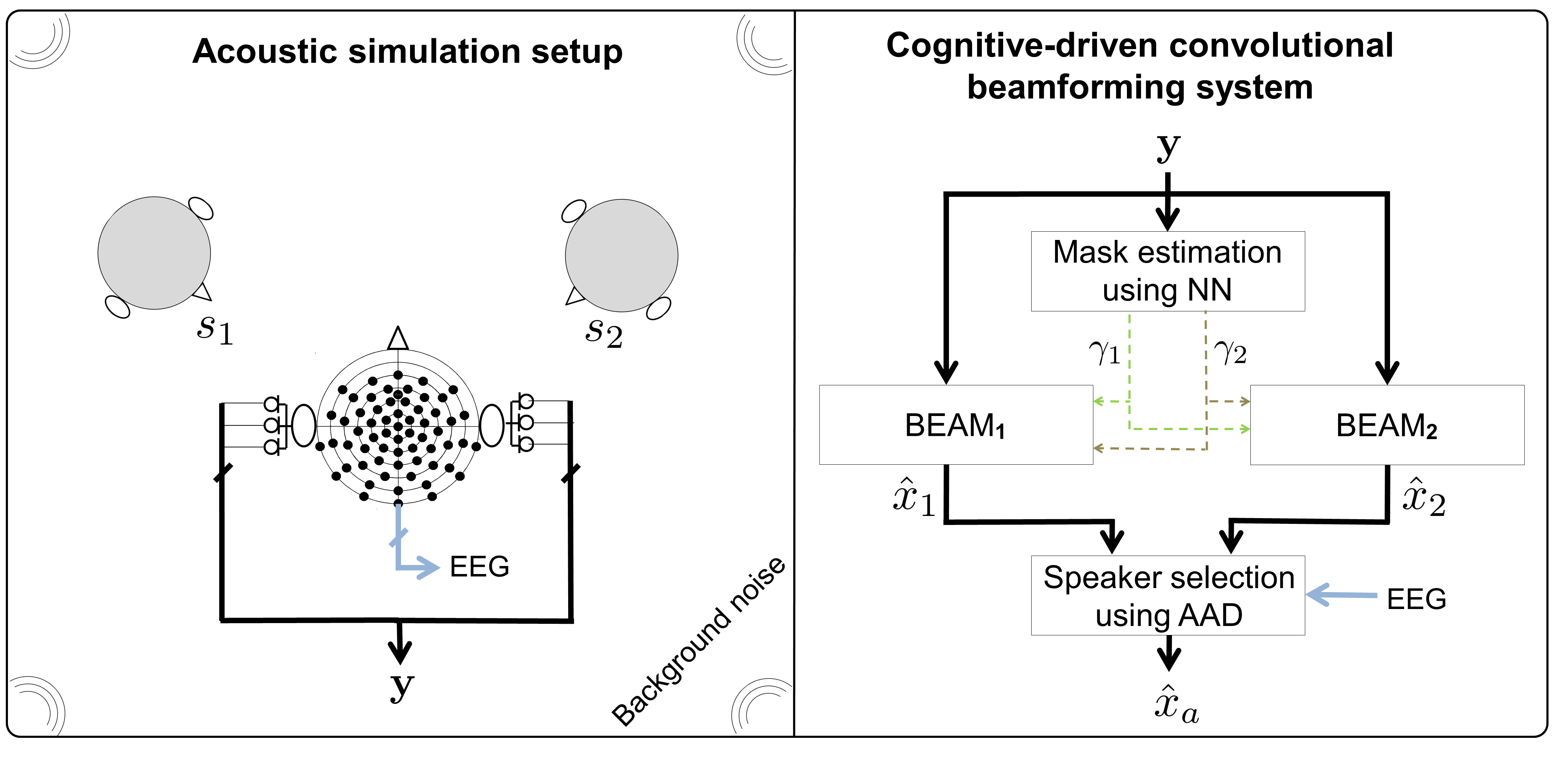}}
  \caption{\small Acoustic simulation setup and block diagram of the proposed cognitive-driven convolutional beamforming system. 
  }
\label{fig:block diagram}
\end{figure}

The proposed system is depicted in Fig. \ref{fig:block diagram} for a scenario with two speakers. 
First, time-frequency masks of both speakers are estimated from the noisy and reverberant microphone signals using  a speaker-independent speech separation neural network. Then, two beamformers are designed to generate reference signals for AAD by enhancing the speech signal of each speaker based on the estimated masks. The AAD method then selects one of the reference signals as the enhanced attended speech signal. For the beamformers we propose to use a recently proposed weighted minimum power distortionless response (wMPDR) convolutional beamformer as it optimally combines dereverberation, noise suppression and interfering speaker suppression \cite{Christoph_2020_ICASSP}. While suppressing the interfering speaker is desired to improve speech intelligibility, keeping the interfering speaker audible is also important to allow the listener to switch attention between speakers. Therefore, we  also  propose  an  extension  of  the  wMPDR  convolutional beamformer  incorporating  an  interference  suppression  constraint,  referred  to  as a  weighted  linearly  constrained  minimum  power (wLCMP)  convolutional beamformer, which allows to control  the level of suppression  of  the interfering  speaker. 

We experimentally compare our proposed method with state-of-the-art cognitive-driven systems based on conventional MPDR, LCMP, MVDR and LCMV beamformers, which are steered based on estimated masks or estimated DOAs. 
The results show that the proposed system outperforms state-of-the-art cognitive-driven systems for dealing with noisy and reverberant speech mixtures and reveal potential future research directions.

%

%
\section{Cognitive-driven convolutional \\ beamformer}
\label{sec: Cognitive-driven convolutional beamformer}

%
\subsection{Signal model}
\label{subsec: configuration and notation}
We consider an acoustic scenario comprising $I$ competing speakers\footnote{It should be noted that we provide a general description of the algorithms for $I$ speakers, but limit our experiments in Section \ref{sec: experimental setup} to two speakers.} with the clean signals denoted as $s_{i}\left[n\right]$, $i=1\;\ldots\;I$ where $n$ is the discrete time index. We consider a binaural hearing aid setup with $M$ microphones.
The $m$-th microphone signal $y_{m}\left[n\right]$ can be decomposed as
\begin{equation}
y_{m}\left[n\right]=\overset{I}{\underset{i=1}{\sum}}x_{i,m}\left[n\right]+v_{m}\left[n\right],\;\;\;\;\;m=1\;\ldots\;M,
\label{eq: mic signal left}
\end{equation}
where $x_{i,m}\left[n\right]$ denotes the reverberant speech component in the $m$-th microphone signal corresponding to speaker $i$ and $v_{m}\left[n\right]$ denotes the background noise component. 
The reverberant speech components $x_{i,m}\left[n\right]$ consist of an anechoic speech component $\ensuremath{x_{i,m}^{an}\left[n\right]}$ (encompassing the head filtering effect), an early reverberation component arriving typically in the order of tens of milliseconds, and a late reverberation component. 
While early reverberation can be beneficial for speech intelligibility, late reverberation is known to have a detrimental effect on speech quality and intelligibility \cite{Warzybok_2013}. 

In the short-time Fourier Transform (STFT) domain, the $M$-dimensional stacked vector of all microphone signals is given by
\begin{equation}
\ensuremath{\mathbf{y}}_{k,f}=\left[Y_{1,k,f}\;\ldots\;Y_{M,k,f}\right]^{T} \in \mathbb{ C}^{M \times 1},
\label{eq: mic signal in STFT}
\end{equation}
where $Y_{m,k,f}$ denotes the STFT coefficient of $y_{m}\left[n\right]$, and $k=1\;\ldots\;K$ and $f=1\;\ldots\;F$ are the frame index and the frequency index, respectively. 

%
\subsection{Mask estimation}
\label{subsec: mask estimation}

The first component of our proposed system is  a separation neural network that estimates time-frequency ideal ratio masks corresponding to each speaker from the reverberant and noisy microphone signals. These masks will be used for beamforming and to generate reference signals for AAD (see Section \ref{subsec: Reference signal generation using beamformers}). 

Several neural network-based speech separation approaches have been proposed, both in frequency-domain and in time-domain \cite{Kolbak_2017_IEEE_ACM_ASLP,Luo_ICASSP_2018}. In this paper we use a BLSTM-based frequency-domain approach~\cite{Kolbak_2017_IEEE_ACM_ASLP} since it trains faster than time-domain approaches such as \cite{Luo_ICASSP_2018}, allowing a faster experimental turnover.


The separation neural network takes the 
STFT coefficients of the $m$-th microphone signal as input features and generates real-valued time-frequency masks, i.e., 
\begin{equation}
    [\bm{\Gamma}_{1,m}\;\ldots\; \bm{\Gamma}_{I+1,m}]=h(\mathbf{Y}_{m}),
    \label{eq: mask estimatin using DNN}
\end{equation}
where the matrix $\mathbf{Y}_{m} \in \mathbb{C}^{K\times F}$ contains all STFT coefficients of the $m$-th microphone signal, $h(\cdot)$ is the separation neural network, and the matrix $\bm{\Gamma}_{i,m}  \in \mathbb{R}^{K\times F}$ for $i=1\;\ldots\;I$, contains the estimated time-frequency masks for speaker $i$. In addition to the time-frequency masks for the speakers, the network also generates a time-frequency mask for the background noise, i.e., $\bm{\Gamma}_{I+1,m}$. 


The separation neural network is trained using permutation invariant training (PIT)~\cite{Kolbak_2017_IEEE_ACM_ASLP} with a scale-dependent SNR loss in the time-domain~\cite{roux2019sdr}. However,  at test time the masks have speaker permutation ambiguity, i.e., it is not known which mask corresponds to which speaker. In addition, the separation neural network in (\ref{eq: mask estimatin using DNN}) operates on each microphone signal independently, which typically causes speaker permutation ambiguities across the microphones. 
To resolve this ambiguity, we align the masks obtained for each microphone based on the least-squares error.
We then average the masks across the microphones to obtain one mask for each speaker, i.e. $\bar{\bm{\Gamma}}_{i}\in\mathbb{\mathbb{R}}^{K\times F}$. The averaged mask $\bar{\bm{\Gamma}}_{i}$ contains the masks $\ensuremath{\gamma}_{i,k,f}$ of the $i$-th speaker for all times frames and frequencies.



%
\subsection{Reference signal generation using beamformers}
\label{subsec: Reference signal generation using beamformers} 
Based on the estimated masks $\bar{\bm{\Gamma}}_{i}$, we design $I$ beamformers to extract each speaker with reduced noise and reverberation from the microphone signals (see $\text{BEAM}_1$  and $\text{BEAM}_2$ in Fig. \ref{fig:block diagram}). The output signals $z_{i,k,f}$ of the beamformers are then transformed to the time-domain as $\hat{x}_{i}\left[n\right]=\textrm{ISTFT}\left( z_{i,k,f}\right) $, where \textrm{ISTFT} denotes the inverse short-time Fourier transform. These time-domain output signals $\hat{x}_{i}\left[n\right]$ will be used as reference signals for AAD.



In this paper we investigate different types of beamformers for generating reference signals, i.e., 
wMPDR and wLCMP convolutional beamformers, and  conventional MPDR and LCMP beamformers, which will be described in detail in Section \ref{sec: Beamforming}. 

%
\subsection{Speaker selection using AAD}
\label{subsec: Speaker selection using AAD}
Based on the reference signals $\hat{x}_{i}\left[n\right]$ generated by the beamformers, the speaker which the listener is attending to is then selected using the EEG-based auditory attention decoding method proposed in \cite{OSullivan2014}. 
First, an estimate of the envelope of the attended speech signal $\hat{e}_{a}\left[l\right]$, with $l$ the sub-sampled time index,  is reconstructed from the EEG signals using a trained spatio-temporal filter. 
Then, the correlation between the reconstructed envelope $\hat{e}_{a}\left[l\right]$ and the envelopes $\hat{e}_{i}\left[l\right]$ of the reference signals $\hat{x}_{i}\left[n\right]$ is computed, i.e., 
\begin{equation}
\rho_{i}=\rho\left(\hat{e}_{i}\left[l\right],\;\;\;\hat{e}_{a}\left[l\right]\right),\;\;\;\;\;i=1\;\ldots\;I,
\label{eq: correlation coff}
\end{equation}
where $\rho(\cdot)$ is the Pearson correlation.
Finally, the attended speech signal  $\hat{x}_{a}\left[n\right]$ is selected as the reference signal yielding the maximum correlation with the reconstructed envelope, i.e.,
\begin{equation}
\hat{x}_{a}\left[n\right]=\hat{x}_{\;\bar{i}}\left[n\right],\;\;\;\;\;\ensuremath{\bar{i}=\underset{i}{\textrm{argmax}}\;\;\rho_{i}}.
\end{equation}

\section{Beamforming}
\label{sec: Beamforming}
In this section, we review the wMPDR convolutional beamformer \cite{Nakatani_EUSIPCO_2019}, present the proposed wLCMP convolutional beamformer, and compare them with the conventional MPDR and LCMP beamformers. Since the beamformer operates for each frequency independently, the frequency index $f$ will be omitted in this section for notational conciseness. 

%
\subsection{Weighted MPDR convolutional beamformer}
\label{subsec: weighted MPDR convolutional beamformer}
The wMPDR convolutional beamformer in \cite{Nakatani_EUSIPCO_2019} aims at 1) suppressing the noise component while preserving the target speech component in one of the microphone signals and 2) suppressing the late reverberation component while preserving the early reverberation component corresponding to the target speaker (i.e., dereverberation). 
The output signal $z_{k}$ of a convolutional beamformer is defined as
\begin{equation}
    z_{k}=\mathbf{\mathbf{\bar{w}}}^{H}\bar{\mathbf{y}}_{k}=\mathbf{w}_{0}^{H}\mathbf{y}_{k}+\overset{L_{w}-1}{\underset{\tau=b}{\sum}}\mathbf{w}_{\tau}^{H}\mathbf{y}_{k-\tau},
    \label{eq: convolutional beamformering}
\end{equation}
%
where $\mathbf{\mathbf{\bar{w}}}=\left[\mathbf{w}_{0}^{T}\;\mathbf{w}_{b}^{T}\ldots\;\mathbf{w}_{L_{w}-1}^{T}\right]^{T}\in \mathbb{ C}^{M\left(L_{w}-b+1\right)\times1}$,  $\ensuremath{\mathbf{\bar{y}}_{k}}=\left[\mathbf{y}_{k}^{T}\;\tilde{\mathbf{y}}_{k}^{T}\right]^{T}\in\mathbb{ C}^{M\left(L_{w}-b+1\right)\times1}$, $\tilde{\mathbf{y}}_{k}$ consists of the observation from $b$ frames in the past until $L_{w}-1$ frames in the past, i.e., $\tilde{\mathbf{y}}_{k}=\left[\mathbf{\mathbf{y}}_{k-b}^{T}\ldots\;\mathbf{y}_{k-L_{w}+1}^{T}\right]^{T}$, and $b$ and $L_{w}$ model the frame delay of the start and  end time of the late reverberation, respectively. 

It has been shown in \cite{Christoph_2020_ICASSP} that the convolutional beamformer $\mathbf{\mathbf{\bar{w}}}$ can be factorized into a dereverberation matrix $\mathbf{G} \in \mathbb{ C}^{M\left(L_{w}-b+1\right)\times M}$ and a beamforming vector $\mathbf{q} \in \mathbb{ C}^{M \times 1}$, i.e., $\mathbf{\mathbf{\bar{w}}}=-\mathbf{Gq}$ with $\mathbf{q}=\mathbf{w}_{0}$. The convolutional beamforming in (\ref{eq: convolutional beamformering}) can hence be written as dereverberation filtering followed by beamforming \cite{Christoph_2020_ICASSP}, i.e.,
\begin{equation}
    \mathbf{d}_{k}=\underset{\textrm{dereverberation}}{\underbrace{\mathbf{y}_{k}-\mathbf{G}^{H}\bar{\mathbf{y}}_{k}}},\;\;\;\;\;z_{k}=\underset{\textrm{beamforming}}{\underbrace{\mathbf{q}^{H}\mathbf{d}_{k}}}.
    \label{eq: dereverberation and beamforming}
\end{equation}

Assuming that the output of the convolutional beamformer $z_{k}$ follows a zero mean complex Gaussian distribution with a time-varying variance \cite{Nakatani_EUSIPCO_2019}, the wMPDR convolutional beamformer is obtained by maximizing an objective function $\mathcal{L}\left(\mathbf{\bar{w}}\right)$, which is derived based on the maximum-likelihood estimation with a target speaker preservation constraint (distortionless constraint), i.e.,
\begin{equation}
    \mathcal{L}\left(\mathbf{\bar{w}}\right)\propto\frac{1}{K}\overset{K}{\underset{k=1}{\sum}}\left(-\ln\left(\lambda_{k}\right)-\frac{\left|z_{k}\right|^{2}}{\lambda_{k}}\right),
    \label{eq: loglikelihood without constraint}
\end{equation}
where $\lambda_{k}$ denotes the time-varying variance of the target speech component (including the early reverberation) and $K$ denotes the number of frames over which the beamformer coefficients are
estimated.  

This optimization problem can be solved in a alternating fashion, by first assuming $\lambda_{k}$ constant and solving for $\mathbf{\bar{w}}$ and then updating $\lambda_{k}$. 
Assuming $\lambda_{k}$ constant, the optimization problem of the wMPDR convolutional beamformer incorporating the target speaker preservation constraint can be written as \cite{Nakatani_EUSIPCO_2019}
\begin{equation}
    \underset{\mathbf{\bar{w}}}{\textrm{max}}\,\,-\mathbf{\bar{w}}^{H}\mathbf{\mathbf{\bar{R}}_{\bar{y}}}\mathbf{\mathbf{\bar{w}}}\;\;\;\;\;\textrm{s.t.}\;\;\;\;\;\underset{\textrm{target}}{\underbrace{\mathbf{\bar{w}}^{H}\bar{\mathbf{a}}=1}},
    \label{eq: loglikelihood with constraint}
\end{equation}
where $\bar{\mathbf{a}}$ denotes the relative early transfer function (RETF) vector corresponding to the target speaker and $\mathbf{\bar{R}}_{\bar{y}}=\frac{1}{K}\sum_{k}\frac{\ensuremath{\mathbf{\overline{y}}_{k}}\ensuremath{\mathbf{\overline{y}}_{k}}^{H}}{\lambda_{k}}$. 
The wMPDR convolutional beamformer solving (\ref{eq: loglikelihood with constraint}) is given by \cite{Christoph_2020_ICASSP}
\begin{equation}
    \mathbf{\mathbf{\bar{w}}}_{\textrm{wMPDR}}=-\mathbf{Gq}_{\textrm{\textrm{wMPDR}}},
    \label{eq: wpe and wMPDR}
\end{equation}
where
\begin{equation}
   \mathbf{G}=\mathbf{R}_{\tilde{y}}^{-1}\mathbf{P}_{\tilde{y}},\;\;\;\;\;\mathbf{q}_{\textrm{wMPDR}}=\frac{\mathbf{R}_{d}^{-1}\mathbf{\bar{\mathbf{a}}}}{\mathbf{\bar{\mathbf{a}}}^{H}\mathbf{R}_{d}^{-1}\mathbf{\bar{\mathbf{a}}}},
    \label{eq: G and q of wpe and wMPDR}
\end{equation}
with $\mathbf{R}_{\tilde{y}}=\frac{1}{K}\sum_{k}\frac{\tilde{\mathbf{y}}_{k}\tilde{\mathbf{y}}_{k}^{H}}{\lambda_{k}}$, $\mathbf{P}_{\tilde{y}}=\frac{1}{K}\sum_{k}\frac{\tilde{\mathbf{y}}_{k}\mathbf{y}_{k}^{H}}{\lambda_{k}}$,  $\mathbf{R}_{d}=\frac{1}{K}\sum_{k}\frac{\mathbf{d}_{k}\mathbf{d}_{k}^{H}}{\lambda_{k}}$. 


To estimate the RETF vector of the target speaker $\bar{\mathbf{a}}$ in (\ref{eq: G and q of wpe and wMPDR}), we use the masks of the target speaker $\gamma_{t,k}$, assuming the target speaker index is $t$. The RETF vector is estimated using the covariance whitening method \cite{Markovich_2009}, i.e.,  

\begin{equation}
    \bar{\mathbf{a}}=\mathbf{R}_{\widetilde{t}+v}\textrm{MaxEig}\left(\mathbf{R}_{\widetilde{t}+v}^{-1}\mathbf{R}_{t}\right),
    \label{eq: RTF for wMPDR}
\end{equation}
    where $\mathbf{R}_{t}=\frac{\sum_{k}\gamma_{t,k}\mathbf{d}_{k}\mathbf{d}_{k}^{H}}{\sum_{k}\gamma_{t,k}}$ is the covariance matrix of the target speaker 
and $\mathbf{R}_{\widetilde{t}+v}=\frac{\sum_{k}\left(1-\gamma_{t,k}\right)\mathbf{d}_{k}\mathbf{d}_{k}^{H}}{\sum_{k}\left(1-\gamma_{t,k}\right)}$ is the covariance matrix of all interfering speakers and background noise. 

The estimation methods discussed in this section are used to iteratively update the output signal of the wMPDR convolutional beamformer. First, the dereverberation filtering in (\ref{eq: dereverberation and beamforming}) is performed using $\mathbf{G}$ in (\ref{eq: G and q of wpe and wMPDR}). Based on the dereverberated signals $\mathbf{d}_{k}$ and the estimated masks $\gamma_{t,k}$, the RETF vector of the target speaker $\bar{\mathbf{a}}$ is updated using (\ref{eq: RTF for wMPDR}) to steer the beamformer $\mathbf{q}_{\textrm{wMPDR}}$ in (\ref{eq: G and q of wpe and wMPDR}). Using the steered beamformer, the output signal $z_{k}$ in (\ref{eq: dereverberation and beamforming}) is obtained. The variance of the target speech component is then updated as $\lambda_{k} = |z_{k}|^2$  for the next iteration.

\subsection{Weighted LCMP convolutional beamformer}
\label{subsec: weighted LCMP convolutional beamformersubsec: weighted MPDR convolutional beamformer}
As an alternative to the wMPDR convolutional beamformer, we propose the wLCMP convolutional beamformer, which allows to control the suppression of the interfering speakers. The wLCMP convolutional beamformer is derived by adding interfering speaker suppression constraints to the optimization problem of the wMPDR convolutional beamformer, i.e., 
\begin{equation}
    \underset{\mathbf{\bar{w}}}{\textrm{max}}\,\,-\mathbf{\bar{w}}^{H}\mathbf{\mathbf{\bar{R}}_{\bar{y}}}\mathbf{\mathbf{\bar{w}}}\;\;\;\;\;\textrm{s.t.}\;\;\;\;\;\underset{\textrm{target}}{\underbrace{\mathbf{\bar{w}}^{H}\mathbf{\bar{a}}=1}},\;\;\;\;\;\underset{\textrm{interference}}{\underbrace{\mathbf{\bar{w}}^{H}\bar{\mathbf{B}}=\mathbf{\boldsymbol{\delta}}}},
    \label{eq: wpe and wLCMP}
\end{equation}
where $\bar{\mathbf{B}}=\left[\bar{\mathbf{b}}_{1}\;\ldots\;\bar{\mathbf{b}}_{U}\right]$ contains the RETF vectors of  $U$ interfering speakers, with $U=I-1$,  and $\boldsymbol{\delta}=\left[\delta_{1}\;\ldots\;\delta_{U}\right]$ controls the amount of suppression of the interfering speakers. This optimization problem is the same as the optimization problem of the conventional LCMP beamformer \cite{ Habets2012}, 
but with different RTF vectors and covariance matrix. Therefore the wLCMP convolutional beamformer can be obtained as 
\begin{equation}
    \mathbf{\mathbf{\bar{w}}}_{\textrm{wLCMP}}=-\mathbf{Gq}_{\textrm{\textrm{wLCMP}}},
     \label{eq: wpe and  wLCMP}
\end{equation}
where the dereverberation matrix $\mathbf{G}$ is obtained as in (\ref{eq: G and q of wpe and wMPDR}) and the beamforming vector $\mathbf{q}_{\textrm{wLCMP}}$ is obtained as in \cite{Habets2012}, i.e.,
\begin{equation}
    \mathbf{q}_{\textrm{wLCMP}}=\mathbf{R}_{d}^{-1}\bar{\mathbf{C}}\left(\bar{\mathbf{C}}^{H}\mathbf{R}_{d}^{-1}\bar{\mathbf{C}}\right)^{-1}\mathbf{p},
    \label{eq: wLCMP}
\end{equation}
with $\ensuremath{\bar{\mathbf{C}}=\left[\bar{\mathbf{a}}\;\;\;\;\;\bar{\mathbf{B}}\right]}$ and $\mathbf{p}=\left[1\;\;\;\;\;\boldsymbol{\delta}\right]^{T}$. 
Setting $\delta_{u}$ to zero in (\ref{eq: wLCMP}) corresponds to a complete suppression of the $u$-th interfering speaker, while $\ensuremath{\delta}>0$ leads to a controlled suppression.

The RETF vector of the target speaker $\bar{\mathbf{a}}$ in (\ref{eq: wLCMP}) is estimated using (\ref{eq: RTF for wMPDR}). The RETF vector of the $u$-th interfering speaker $\bar{\mathbf{b}}_{u}$  is estimated as 
\begin{equation}
    \bar{\mathbf{b}}_{u}=\mathbf{R}_{\widetilde{u}+v}\textrm{MaxEig}\left(\mathbf{R}_{\widetilde{u}+v}^{-1}\mathbf{R}_{u}\right)
\end{equation}
where $\mathbf{R}_{u}=\frac{\sum_{k}\gamma_{u,k}\mathbf{d}_{k}\mathbf{d}_{k}^{H}}{\sum_{k}\gamma_{u,k}}$ is the covariance matrix of the $u$-th interfering speaker and $\mathbf{R}_{\widetilde{u}+v}=\frac{\sum_{k}\left(1-\gamma_{u,k}\right)\mathbf{d}_{k}\mathbf{d}_{k}^{H}}{\sum_{k}\left(1-\gamma_{u,k}\right)}$.

The output signal of the wLCMP convolutional beamformer is iteratively updated similarly as for the wMPDR convolutional beamformer.

\subsection{Relation with conventional MPDR and LCMP beamformers}
\label{subsec: MPDR and LCMP}

The conventional MPDR beamformer aims at minimizing the PSD of the output signal while preserving the reverberant target speech component in one of the microphone signals \cite{ Simon2015}. The MPDR beamformer is given by
\begin{equation}
    \mathbf{w}_{{\scriptstyle \text{MPDR}}}=\frac{\mathbf{\mathbf{R}}_{y}^{-1}\mathbf{a}}{\mathbf{a}^{H}\mathbf{\mathbf{R}}_{y}^{-1}\mathbf{a}},
    \label{eq: MPDR}
\end{equation}
where $\mathbf{\mathbf{R}}_{y}=\frac{1}{K}\sum_{k}\mathbf{y}_{k}\mathbf{y}_{k}^{H}$ and $\mathbf{a}$ denotes the reverberant RTF vector corresponding to the target speaker.
The MPDR beamformer in (\ref{eq: MPDR}) is similar to the convolutional wMPDR beamformer in (\ref{eq: G and q of wpe and wMPDR}) 
except that the covariance matrix $\mathbf{\mathbf{R}}_{y}$ and the RTF vector $\mathbf{a}$ are estimated using the microphone signals $\mathbf{y}_{k}$ instead of the dereverberated microphone signals $\mathbf{d}_{k}$. In addition, the MPDR beamformer is obtained using a non-iterative optimization procedure compared to the wMPDR convolutional beamformer.

A similar relation exists between the conventional LMCP beamformer incorporating interfering speaker suppression constraints and the wLMCP convolutional beamformer in (\ref{eq: wLCMP}). The conventional LCMP beamformer is given by  \cite{ Habets2012}
\begin{equation}
    \mathbf{w}_{{\scriptstyle \text{LCMP}}}=\mathbf{\mathbf{R}}_{y}^{-1}{\mathbf{C}}\left({\mathbf{C}}^{H}\mathbf{\mathbf{R}}_{y}^{-1}{\mathbf{C}}\right)^{-1}\mathbf{p},
    \label{eq: LCMP}
\end{equation}
with ${\mathbf{C}}=\left[\mathbf{a}\;\;\;\;\mathbf{B}\right]$ and $\mathbf{B}=\left[\mathbf{b}_{1}\;\ldots\;\mathbf{b}_{U}\right]$ containing the reverberant RTF vectors of $U$ interfering speakers. 

The output signals of the MPDR and the LCMP beamformer are obtained as
\begin{equation}
    \ensuremath{z_{k}=\mathbf{w}_{\left\{ {\scriptstyle \text{MPDR, LCMP}}\right\} }^{H}\ensuremath{\mathbf{y}_{k}}}.
\end{equation}
These output signals are obviously computed without involving a dereverberation step compared to the output signals of wMPDR and wLCMP convolutional beamformers in (\ref{eq: convolutional beamformering}).

\section{Experimental setup}
\label{sec: experimental setup}

%
\subsection{Acoustic simulation setup}
\label{subsec: Acoustic simulation setup}
In the experimental evaluation we consider two competing speakers, i.e., $I=2$.  Two German audio stories, uttered by two different male speakers, were used as the clean speech signals $s_{1}\left[n\right]$ and $s_{2}\left[n\right]$. Speech pauses that exceeded $0.5$ s were shortened to $0.5$ s, resulting in two highly overlapping (competing) audio stories. 
The hearing aid microphone signals $y_{m}\left[n\right]$ were generated at a sampling frequency of $16$ kHz by convolving the clean speech signals with non-individualized measured binaural impulse responses (anechoic or reverberant) from \cite{Kayser2009}, and adding diffuse babble noise, simulated according to \cite{Habets2008}. The hearing aid setup in \cite{Kayser2009} consisted of two hearing aids, each equipped with three microphones ($M=6$), mounted on a dummy head. 
The left and the right competing speaker were simulated at $\theta_{1}=-45^{\circ}$ and $\theta_{2}=45^{\circ}$. 
We consider three acoustic conditions, i.e., an anechoic-noisy condition with an average frequency-weighted segmental SNR ($\mathrm{fwSSNR}$) of $2.9$ dB, a reverberant condition (reverberation time $T_{60}\approx0.5$ s) with an average $\mathrm{fwSSNR}$ of $3.5$ dB, and a reverberant-noisy condition with an average $\mathrm{fwSSNR}$ of $0.5$ dB. 
The average $\mathrm{fwSSNR}$ is computed by averaging the highest $\textrm{fwSSNR}$ corresponding to speaker $1$ and to speaker $2$ among the microphone signals. 
The reference signals used to compute the $\textrm{fwSSNR}$ are the anechoic speech signals $\ensuremath{x_{i,m}^{an}\left[n\right]}$ of the speakers at the first microphone of the hearing aid located at the same side of each speaker.


%
\subsection{Mask estimation }
\label{subsec: setup-mask estimation algorithm}
The mask estimation neural network consisted of 3 BLSTM layers of 896 units.
The network was trained on simulated noisy and reverberant mixtures obtained by mixing Librispeech~\cite{7178964} utterances convolved with room impulse responses generated with the image method for reverberation times between $0.2$ s and $0.6$ s, and adding babble noise at $\mathrm{SNRs}$  between 5 and 15 dB. The number of training mixtures was 50k.
Note that there is a large mismatch between the training and the testing condition with respect to reverberation, background noise and head shadow effect, and also a large linguistic dissimilarity, as Librispeech consists of English read speech but the test data consists of German audio stories.


%
\subsection{Beamforming}
\label{subsec: setup-RTF vector estimation algorithm}
All considered beamformers were implemented using a weighted overlap-add (WOLA) framework with an STFT frame length $\mathrm{FL}=512$, an overlap of $75\%$ between successive frames and a Hann window. 
For the wMPDR and wLCMP convolutional beamformers, the frame delay $b$ was set to $4$ and the length of the dereverberation filter was set to $L_{w}=20$, $16$ and $8$ for frequency ranges $0-0.8$kHz, $0.8-1.5$kHz and $1.5-3$kHz, respectively.  The variance of the target speech component was initialized as $\lambda_{k}=\left\Vert \mathbf{y}_{k}\right\Vert ^{2}$. For the wLCMP convolutional beamformer and the LCMP beamformer, we set the interference suppression parameter to $\delta=0.1$ to partially suppress the unattended speaker. The outputs signal of the wMPDR and wLCMP convolutional beamformers were obtained with $10$ iterations.

To investigate the impact of mask estimation errors on the speech enhancement performance of the proposed system, we consider oracle ideal ratio masks (oMASK) and estimated ideal ratio masks (eMASK), obtained by the mask estimation neural network in (\ref{eq: mask estimatin using DNN}).

We also compare our proposed system with  a state-of-the-art  cognitive-driven system proposed in \cite{Aroudi_2020_cognitive-driven_beamforming_ASLP}, which uses either a conventional MVDR beamformer or a conventional LCMV beamformer to generate reference signals. 
Contrary to the MPDR and LCMP beamformers described in Section \ref{subsec: MPDR and LCMP}, these MVDR and LCMV beamformers use a diffuse noise covariance matrix instead of $\mathbf{\mathbf{R}}_{y}$ and are steered using estimated anechoic RTF vectors (based on estimated DOAs of both speakers) instead of estimated reverberant RTF vectors. 
For the LCMV beamformer, the interference suppression parameter was set to $\delta=0.1$. Similarly as in \cite{Aroudi_2020_cognitive-driven_beamforming_ASLP}, the DOAs of both speakers were estimated using a classification-based method \cite{Hendrik_2014_DOA} and the anechoic RTF vectors corresponding to the estimated DOAs were selected from a database of (measured) prototype RTF vectors \cite{Kayser2009}. 

%
\subsection{Speaker selection using AAD}
\label{subsec: setup-speaker selection using AAD}
We used EEG responses recorded for 16 native German-speaking participants, where $8$ participants were instructed to attend to the left speaker and $8$ participants to the right speaker. 
See \cite{Aroudi_2020_cognitive-driven_beamforming_ASLP} for details about the EEG recording and the AAD training and decoding configuration. 

For the AAD training and decoding steps (see Section \ref{subsec: Speaker selection using AAD}), 
the EEG recordings were split into $30$-second trials, resulting in $40$ trials for the anechoic-noisy condition as well as for the reverberant-noisy condition, and $20$ trials for the reverberant condition. 
Each participant's own data were used for training the spatio-temporal filter used for reconstructing the speech envelope $\hat{e}_{a}[l]$ from the EEG data.

\subsection{Performance measures}
We evaluate the cognitive-driven beamformers both in terms of AAD and speech enhancement performance.
To evaluate the AAD performance, a trial is considered to be correctly decoded if the $\textrm{fwSSNR}$ corresponding to the selected beamformer output signal $\hat{x}_{a}\left[n\right]$ (as the attended speech signal) is larger than the $\textrm{fwSSNR}$ corresponding to the discarded beamformer output signal. To compute $\textrm{fwSSNR}$, the anechoic speech component $\ensuremath{x_{a,m}^{an}\left[n\right]}$ of the attended speaker in the first microphone signal of the hearing aid at the side of the attended speaker was used as the $\textrm{fwSSNR}$ reference signal. 
The AAD performance is then computed by averaging the percentage of correctly decoded trials over all considered trials and all participants. 

The speech enhancement performance of the complete proposed system is evaluated in terms of the $\textrm{fwSSNR}$ improvement ($\Delta\textrm{fwSSNR}$) using the same reference signals as used for AAD performance evaluation. The input $\textrm{fwSSNR}$ is defined as the highest $\textrm{fwSSNR}$ among the microphone signals. 
The output $\textrm{fwSSNR}$ is defined as the $\textrm{fwSSNR}$ of the selected beamformer output signals $\hat{x}_{a}\left[n\right]$. 




To investigate the impact of the errors of speaker selection using AAD on the speech enhancement performance of the complete proposed system, we will consider oracle AAD (oAAD) where the attended speech signal $\hat{x}_{a}\left[n\right]$ is determined based on the highest  $\Delta\textrm{fwSSNR}$ among the output signals of $\text{BEAM}_1$  and $\text{BEAM}_2$, and estimated AAD (eAAD) where $\hat{x}_{a}\left[n\right]$ is determined based on the highest Pearson correlation coefficients as described in Section\ref{subsec: Speaker selection using AAD}.

%
\section{Experimental results}
\label{sec: experimental results}
In this section, we evaluate the AAD performance and the speech enhancement performance of the proposed cognitive-driven convolutional beamforming system. 
In Section \ref{subsec: decoding performance} we investigate the impact of mask estimation errors on the AAD performance. In Section \ref{subsec: speech enhancement performance}, we investigate the impact of AAD errors on the speech enhancement performance.

%
\begin{figure}[t]
 \centering
  \centerline{\includegraphics[width=7cm,height=4cm]{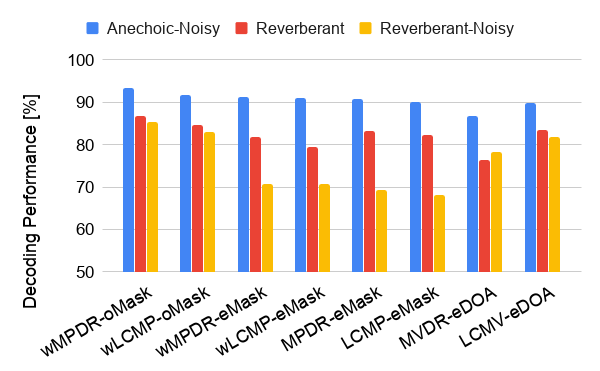}}
 \caption{\small Average auditory attention decoding performance for the anechoic-noisy, reverberant and reverberant-noisy conditions for the different considered beamformers. The upper boundary of the confidence interval corresponding to chance level for the anechoic-noisy, reverberant and reverberant-noisy conditions are $61.39\%$, $66.19\%$, $61.39\%$, respectively, computed based on a binomial test at the $5$\% significance level.} 
\label{fig:decoding performance}
\end{figure}

%
\subsection{Auditory attention decoding performance}
\label{subsec: decoding performance}
Figure \ref{fig:decoding performance} depicts the average AAD performance for the anechoic-noisy, the reverberant and the reverberant-noisy condition, when using the output signals of the wMPDR or wLCMP convolutional beamformer, the MPDR or LCMP beamformer and the MVDR or LCMV beamformer as reference signals for decoding. 
We observe that all considered beamformers yield a AAD performance that is significantly larger than chance levels. 
For all considered acoustic conditions the wMPDR convolutional beamformer and the wLCMP convolutional beamformer using the oracle masks (wMPDR-oMASK and wLCMP-oMASK) yield the highest AAD performance,  
showing the potential of using convolutional beamformers for AAD.

When using estimated masks instead of oracle masks for the convolutional beamformers (wMPDR-eMASK and wLCMP-eMASK) the AAD performance decreases,  especially in the reverberant-noisy condition. 
In the reverberant-noisy condition, the MVDR and LCMV beamformers using anechoic RTF vectors based on  estimated DOAs (MVDR-eDOA and LCMV-eDOA) yield a larger average AAD performance than the beamformers using reverberant RTF vectors based on the estimated masks. 
This suggests that in order to improve the AAD performance, a better estimation of RTF vectors is required, e.g., based on prototype RTF vectors or neural networks that are more robust to background noise and reverberation. 

\begin{figure}[t]
\centering
\begin{subfigure}[b]{.50\linewidth}
  \centerline{\includegraphics[width=7cm,height=4.2cm]{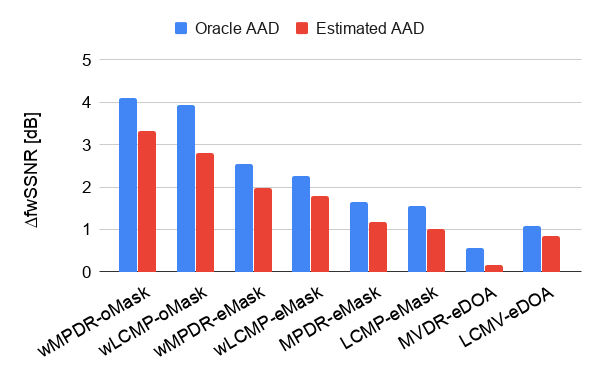}}
\centerline{(a)}\medskip
\end{subfigure}

\begin{subfigure}[b]{.50\linewidth}
  \centerline{\includegraphics[width=7cm,height=4.2cm]{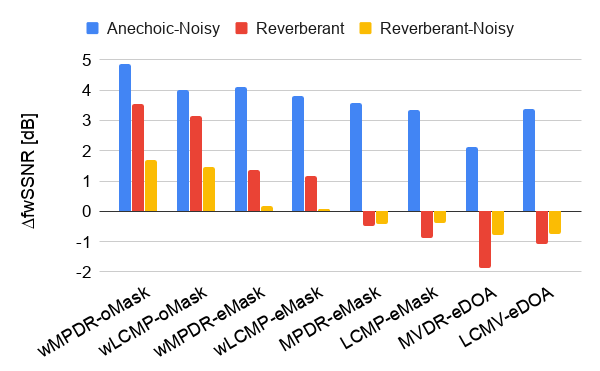}}
\centerline{(b)}\medskip
\end{subfigure}
\caption{\small $\textrm{fwSSNR}$ improvement (a) averaged over all considered acoustic conditions when using oracle AAD and estimated AAD (b) for the anechoic-noisy, reverberant and reverberant-noisy conditions when using estimated AAD. The input $\textrm{fwSSNR}$ averaged over all considered acoustic conditions is $2.06$dB and the input $\textrm{fwSSNRs}$ for the anechoic-noisy, reverberant and reverberant-noisy conditions are $2.9$dB, $3.5$dB, $0.5$dB, respectively.}
\label{fig: fwSSNR}
\end{figure}

%
\subsection{Speech enhancement performance}
\label{subsec: speech enhancement performance}
Figure \ref{fig: fwSSNR}a depicts the $\textrm{fwSSNR}$ improvement of the complete proposed
system averaged over all considered acoustic conditions, either using oracle AAD or estimated AAD. 
It can be observed that the convolutional beamformers outperform all other considered beamformers for both oracle and estimated AAD. When using estimated AAD instead of oracle AAD, for all considered beamformers the $\textrm{fwSSNR}$ improvement decreases by $0.2$--$1.1$ dB, showing the sensitivity to AAD errors. 
Nevertheless, the $\textrm{fwSSNR}$ improvement of the convolutional beamformers is about  $1.6$--$1.8$ dB larger than the state-of-the-art MVDR and LCMV beamformers using estimated DOAs.



Figure \ref{fig: fwSSNR}b depicts the $\textrm{fwSSNR}$ improvement of the complete proposed system for the anechoic-noisy, reverberant and reverberant-noisy conditions when using estimated AAD. It can be observed that all beamformers yield a significant $\textrm{fwSSNR}$ improvement for the anechoic-noisy condition. However, for the reverberant condition the systems using  conventional beamformers (MPDR-eMask, LCMP-eMask, MVDR-eDOA, LCMV-eDOA) tend to degrade the $\textrm{fwSSNR}$, whereas only the proposed system using convolutional beamformers (wMPDR-eMask, wLCMP-eMask) provides a  $\textrm{fwSSNR}$ improvement, showing the influence of dereverberation. It should be noted that the considered reverberant-noisy condition with an interfering speaker is an extremely adverse condition with babble noise at a signal-to-interference-plus-noise ratio (SINR) of $0.3$ dB and a reverberation time of $0.5$ s, which makes it very challenging for speech enhancement.

\subsection{Discussion}



The experimental results show that for the considered acoustic setup the AAD performance and the $\textrm{fwSSNR}$ improvement of the proposed cognitive-driven speech enhancement system using convolutional beamformers are sensitive to mask estimation errors, particularly for the reverberant and reverberant-noisy conditions. The mask estimation errors can be mainly attributed to the linguistic dissimilarity of training and testing conditions of the neural-network-based mask estimation algorithm and also the intrinsic difficulty of separating out two competing speakers with the same gender in the reverberant-noisy condition.

The results show that the wMPDR convolutional beamformer yields a larger $\textrm{fwSSNR}$ improvement than the wLCMP convolutional beamformer. Although the wMPDR convolutional beamformer can strongly suppress the interfering speaker, it may deprive the listener from the ability to switch attention between the speakers. In contrast, the wLCMP convolutional beamformer is able to both control the interfering speaker suppression as well as yield  a considerable  $\textrm{fwSSNR}$ improvement. 

Lastly, the results show that the convolutional beamformers (wLCMP-eMASK and wMPDR-eMASK) yield the highest $\textrm{fwSSNR}$ improvement for all considered acoustic conditions, whereas the conventional LCMV beamformer (LCMV-eDOA) yields the highest AAD performance in the reverberant and reverberant-noisy conditions.  Future work could therefore investigate the potential of combining the convolutional and the conventional beamformers to improve both the decoding and the speech enhancement performance.


%
\section{Conclusion}
\label{sec: Discussion}
In this paper, we proposed a cognitive-driven speech enhancement system which combines  neural-network-based  mask estimation, convolutional beamformers and AAD. We considered the wMPDR convolutional beamformer, which jointly enhances the attended speaker and suppresses the unattended speaker, reverberation and background noise. In addition, we proposed a wLCMP convolutional beamformer which enables to control the amount of suppression for the unattended speaker. 
The experimental results showed that the proposed system using convolutional beamformers is able to considerably improve the $\mathrm{fwSSNR}$ both for noisy and reverberant conditions compared to state-of-the-art cognitive-driven speech enhancement systems.     


\bibliographystyle{IEEEtran}
\bibliography{mainRefs_3}

\end{document}